\documentclass[fleqn,usenatbib]{mnras}

\usepackage{color}
\usepackage{amssymb}
\usepackage[pdftex]{graphicx}
\usepackage{epstopdf}
\usepackage{amssymb}
\usepackage{aecompl}
\usepackage{soul}
\usepackage{amsmath}
\usepackage{caption}
\usepackage{float}

\usepackage[titletoc]{appendix}

\usepackage{eso-pic}

\title[Measuring dust with SED fitting]
{Measuring the dust attenuation law of galaxies using photometric data}

\author[Meldorf, Palmese \& Salim]{
\parbox{\textwidth}{
\Large
C. Meldorf$^{1}$, 
A.~Palmese$^{2,3}$\thanks{E-mail: \url{palmese@cmu.edu}}\thanks{NASA Einstein Fellow}
S.~Salim$^{4}$
}
\\
$^{1}$ Department of Astronomy and Astrophysics, University of Chicago, Chicago, IL 60637, USA\\
$^{2}$ McWilliams Center for Cosmology, Department of Physics, Carnegie Mellon University, Pittsburgh, PA 15213, USA\\
$^{3}$ Department of Physics, University of California, Berkeley, 366 Physics North MC 7300, Berkeley, CA 94720, USA\\
$^{4}$ Department of Astronomy, Indiana University, Bloomington, IN 47405, USA\\
}

\begin{document}
\maketitle

\pubyear{2022}

\label{firstpage}

\begin{abstract}
Fitting model spectral energy distributions (SED) to galaxy photometric data is a widely used method to recover galaxy parameters from galaxy surveys. However, the parameter space used to describe galaxies is wide and interdependent, and distinctions between real and spurious correlations that are found between these parameters can be difficult to discern. In this work, we use the SED fitting code \texttt{BAGPIPES} to investigate degeneracies between galaxy parameters and the effect of the choice of different sets of photometric bands. In particular, we focus on optical to infrared wavelength coverage, and on two parameters describing the galaxies' dust attenuation law: $A_V$ and $\delta$, which characterize dust column density and the slope of a flexible dust attenuation law, respectively. We demonstrate that 1) a degeneracy between the residual (the difference between truth and recovered value) $A_V$ and star formation rate exists, but this is lifted when WISE bands are included; 
2) \texttt{BAGPIPES} is able to accurately recover the input $A_V$ and $\delta$ distributions and relations (differences in slope of less than 1.7$\sigma$ for a flat relation, less than 1.2$\sigma$ for an observationally-motivated relation from \citealt{Salim18}) and is not introducing spurious correlations between these parameters. Our findings suggest that the information needed to constrain $A_V$ and $\delta$ well enough individually exists in the data, especially when IR is added.
This indicates that recent works finding a correlation between $A_V$ and $\delta$ are not being misled by fitting degeneracies from their SED fitting code.

\noindent
\end{abstract}


\section{Introduction}

Galaxies are incredibly complicated systems, where their defining properties are highly dependent on their evolution and environment. In addition, the information needed to describe a galaxy is quite varied, typically a model needs to account for their stellar mass, rate of star formation, dust content, metallicity, star formation history, and redshift (see \citep{ConroyReview} for a review). All of this information about the galaxy is compressed into the observational signal we observe, its electromagnetic radiation in a form of a Spectral Energy Distribution (SED). Comparing the observed SED to synthetic models is an extremely widespread way of deriving the physical properties of galaxies. 

Several works have explored the parameter degeneracies and systematics associated to SED fitting. For instance, studies have revealed the inherent relationships between choice of star formation history (SFH) parameterization (or lack thereof, non-parameteric models have proven promising, see \citealt{Leja_2019, Lower_2020}) and recovered stellar mass  \citep{Lower_2020} or age \citep{simha2014parametrising}, treatement of metallicity and stellar mass \citep{Mitchell}, and the degeneracy between age and metallicity \citep{AgeMetal}. 
Other studies have investigated the effects of galaxy dynamics on SED fitting, considering phenomena such as quenching \citep{Ciesla}, morphology \citep{WuytsMergers} or past mergers \citep{Zine_2022}. Systematic uncertainties have also been shown to enter results as a function of redshift when attempting to determine stellar masses \citep{Vanderwel, Paulino}.
Despite this, little exploration has been performed with a varying dust attenuation law, though studies have been done which found that dust's effect on other galaxy parameters is non-negligible \citep{LoFaro, Leja_2018, Lower_2022_dust}. Due to the attenuation law parameters introducing extra degeneracies with other physical properties and with each other \citep{Qin_2022}, most analyses tend to assume a fixed slope for it. 

The {assumption of a fixed slope} is not the case for recent works that have attempted to recover the dust law for Supernovae (SN) host galaxies \citep{MassStepDust, Dixon22}. These works in particular sought to resolve previously unexplained correlations between the host galaxy mass and Hubble diagram residuals, called the mass step. As physical parameters such as dust and stellar mass can or are already used through the aforementioned correlations to improve the standardization of supernovae \citep{Sullivan_2011, Betoule_2014}, a biased estimate of these parameters could lead to systematics in the resulting cosmological measurements. For example, due to the dimming caused by dust, failing to account for dust can cause an overestimation of luminosity distances to these SN. Hence, systematic errors may be introduced into measurements of parameters calculated from the standard candle relationship of Type Ia SNe, such as the dark energy equation of state parameter, $w$, or the matter density of the universe $\Omega_m$, if galaxy properties are not properly estimated \citep{Paulino}.

In this work, we focus specifically on the dust content of galaxies. The effect of dust on the light we observe is twofold. The first effect is extinction: light from the galaxy is absorbed and reddened through interactions with dust particles. 
The second effect is attenuation. Attenuation is a broader phenomenon which includes extinction effects, but also accounts for the redirection of light by scattering off dust. This has the dual effect of reflecting light that would have missed the observer had there been no dust into the path of the observer and vice versa. We describe these phenomena with attenuation laws or curves. These curves represent the ratio between luminosity emitted (no dust interaction) and received (affected by dust) as a function of wavelength.

While this attenuation curve is conveniently specified by a small number of parameters, fitting these parameters accurately can be challenging, as the dust parameters $A_V$ and $\delta$ tend to be degenerate with other properties such as star formation rate, and with each other by definition. Many recent works have found that a higher value for $A_V$ correlates with a flatter attenuation curve, i.e. a higher value of $\delta$\citep{refId0, Kriek_2013, Salmon_2016, Leja_2017, Salim18, Decleir2019, Battisti_2020, Boquien2022}.
However, some recent works \citep{Qin_2022} have claimed that said $A_V - \delta$ relationship could be driven by a degeneracy between the fitting parameters.
We attempt to reproduce the results of \cite{Qin_2022} to determine if the $A_V - \delta$ correlation is driven by fitting degeneracies.

In this work, we seek to explore the effect that the SED fitting process has on recovering specific parameters. We first test the reliability of SED fitting for parameters of interest by considering the distributions of residuals between truth values (the values the models were generated with) and recovered estimations of these parameters. We then specifically analyze the distributions of dust parameter residuals, searching for correlations between the main dust parameters $A_V$ and $\delta$, other parameters, and their respective errors or bias distribution scatter. We finally turn to the analysis performed in \cite{Qin_2022} and attempt to reproduce their results. We extend their analysis to a data-driven $A_V - \delta$ input distribution. Throughout this paper, we will note how different combinations of input bands used for data affect the results. In particular, we compare different combinations of the Dark Energy Camera (DECam, \citealt{decam}) $ugriz$ bands, Visible and Infrared Survey Telescope for Astronomy (VISTA; \citealt{vista}) $JHK_s$ bands, and Wide-Field Infrared Survey Explorer (WISE, \cite{WISEBands}) bands. We only focus on photometric data, although assume that the redshift is  known from spectroscopy (similarly to the analysis in \citealt{MassStepDust}).  
This paper is organized as follows. In \S \ref{methods} we describe our SED-fitting code and the model galaxies used in our analyses. In \S \ref{sec:results} we present our results and discuss, and conclude in \S \ref{sec:conclusions}. All error bars are 1$\sigma$ unless otherwise stated.

\section{Methods}
\label{methods}
Our model SEDs are both created and fitted using the Bayesian Analysis of Galaxies for Physical Inference and Parameter EStimation (\texttt{BAGPIPES}; \citealt{Carnall_2018}) software. \texttt{BAGPIPES} is a fully Bayesian spectral-fitting code used to estimate galaxy properties from photometric and spectroscopic data. \texttt{BAGPIPES} also allows us to reverse this process: estimating the SED for a galaxy with user-given parameters and the measured fluxes in specific photometric bands of the theoretical SED. \texttt{BAGPIPES} allows us to analyze the effect sophisticated dust attenuation curves as well as several parametric SFHs have on measured SEDs. \texttt{BAGPIPES} utilizes the stellar population models derived in \citet{Bruzual_2003} and relies on the \texttt{MultiNest} nested sampling algorithm (\citet{Feroz_2008}; \citet{Feroz2009}, 2013), specifically implemented through the PyMultiNest interface \citep{PyMultiNest}, to obtain the posterior distributions for desired parameters based on SFH and dust models, prior distributions and observational data provided by the user.

\subsection{Attenuation Curves}\label{sec:AttCurves}
Parametric models of attenuation laws come in a variety of types; 
this work considers the \cite{Noll2009} modification of the Calzetti law as formulated by \cite{Salim18}. This flexible attenuation law can take a variety of shapes and is governed by three
parameters: $A_V$, $\delta$ and $B$.
$A_V$ is the dust attenuation in the V band (at $\lambda \simeq 5500 \mathrm{\AA}$). It can be considered as a normalization constant for the attenuation curve, and is proportional to dust column density along the line of sight from object to observer. A higher $A_V$ indicates a dustier environment, and will result in a higher attenuation curve at all wavelengths. $\delta$ is a parameter that governs the difference between the Calzetti and Salim attenuation curves, at $\delta = 0$ they are equivalent. In addition, $\delta$ is directly related to the parameter $R_V$ via the relationship:
\begin{equation}
    R_{V}=\frac{R_{V, \mathrm{Cal}}}{\left(R_{V, \mathrm{Cal}}+1\right)(4400 / 5500)^{\delta}-R_{V, \mathrm{Cal}}} \, , \label{eq:rv}
\end{equation} 
where $R_{V, \mathrm{Cal}} = 4.05$.
The final parameter, $B$, represents the strength of a ``bump" in total attenuation which peaks at a wavelength of 2175 $\mathrm{\AA}$. In previous works \citep{MassStepDust} and in tests described below, we find that letting this parameter vary freely has minimal effect on the recovery of physical parameters, and we therefore choose to set it to $0$ for most of this analysis. 

\subsection{Simulations}\label{sec:sims}

Each model galaxy is given a unique combination of input parameters that govern its dust attenuation curve, SFH and redshift. We simulate galaxies in two different ways, which are used in different parts of the analysis. The first simulation set (hereinafter \texttt{Sim1}) is made in a grid of parameter values, while the second set (hereinafter \texttt{Sim2}) follows a continuous, realistic distribution of parameters.

For \texttt{Sim1}, each parameter and its values are given in Table \ref{table:params}. The dust parameters considered are those given in Section \S \ref{sec:AttCurves}, and truth values are selected to be evenly spaced {between the minimum and maximum value we consider.} This lets us explore the effect of the SED fitting as a function of the input parameters more broadly than we would with a more realistic distribution of galaxies where the bulk of the objects then to cluster around specific parameters values. Hence, $A_V$ has input values of $0.100$, $0.525$, $0.950$, $1.375$, and $1.800$, and $\delta$ is given $-1.400$, $-0.975$, $-0.550$, $-0.125$, $0.300$.
The SFH of each galaxy is parameterized as a log-normal function, which was first utilised in \citet{gladders} and revised in \citet{simha2014parametrising}:
\begin{equation}
{\rm SFR}(t) \propto \frac{1}{t} {\rm exp}\Big[ -\frac{({\rm ln}(t)-T_{0,l})^2}{2\tau_l^2} \Big]\, , 
\end{equation}
where {$t$ is the age of the universe,} SFR is the star formation rate, and $\tau_l$ and $T_{0,l}$ are free parameters. We follow \citet{Diemer_2017} and \citet{Carnall_2019} in redefining these parameters in terms of the more intuitive $t_{\rm max}$ (time at which SFR is maximal) and $\sigma_{\rm SFH}$ (width of the SFH peak at half maximum height) as follows:
\begin{equation}
     t_{\rm max} = e^{T_{0, l}-\tau_{l}^2} \, ,
\end{equation}
and
\begin{equation}
   \sigma_{\rm SFH} = 2t_{\rm max}\hspace{5 pt}{\rm sinh}(\sqrt{2{\rm ln}(2)}\tau_{l}) \, .
\end{equation}

We again pick our input values for these parameters to be uniformly spaced throughout the range of possible values, though for these SFH parameters we additionally choose to avoid including values that would be significantly longer than the age of the universe. Thus, $t_{\rm max}$ has input values $2$, $4.66$, $7.33$, and $10$ Gyrs, while $\sigma_{\rm SFR}$ is given $4$, $7.33$, $10.66$, $14$ Gyrs as input values. The redshift values chosen are $0.2$, $0.4$, $0.6$, $0.8$, so as to match most of the galaxies with spectroscopic redshifts in e.g. the {Dark Energy Survey Supernova} (DES SN) program. For the total mass formed, we again pick evenly spaced values within the prior range, though for this parameter they are spaced logarithmically, giving log$_{10}$(M$_{\mathrm{form}}$/M$_\odot$) = $8.5$, $9.5$, $10.5$, and $11.5$. The initial mass function used by \texttt{BAGPIPES} is given in \cite{IMF}. 
We similarly choose values for the metallicity of the galaxy, $Z$, to be geometrically uniformly spaced within the prior, specifically $0.5$, $1.0$, and $2.0$ times the solar metallicity (Z$_{\odot}$).

In order to create our models, we consider every possible combination of all parameters, leading to a total number of models that is the product of the number of values considered for each parameter. For the values in Table \ref{table:params}, this results in 19,200 unique models {in total.} 

Entering these parameters into \texttt{BAGPIPES} yields values for photometric measurements of each model galaxy, specifically we produce photometric data in $ugriz$ bands of the Dark Energy Camera (DECam; \citealt{flaugher}), and the Visible and Infrared Survey Telescope for Astronomy (VISTA; \citealt{vista}) $JHK_s$ bands.  We choose to include the same bands used in the DES deep fields, as in the DES SN host studies of \citet{MassStepDust}, and also include the WISE 1,2,3 and 4 bands, since they are available over the entire sky, and specifically in combination with $grz$ optical bands in the Dark Energy Spectroscopic Instrument (DESI) Legacy Survey \citep{legacysurvey}.

In order to simulate measurement noise, we assume a signal-to-noise ratio of ten and assign to each data point an error of its flux value divided by ten. This is a conservative choice if we want to use our findings for the DES deep fields or Rubin LSST, where the optical bands reach a greater signal to noise. We then pick a random value drawn from a normal distribution centered at the true flux  with a standard deviation of the error to each photometric data point. A similar process is applied to the redshift measurements, where we take the error to be 0.001.

Whereas in \texttt{Sim1} we allow for every possible combination of every parameter, in \texttt{Sim2} we specifically select for realistic combinations that could be seen in galaxies. We still use \texttt{Sim1}, although it is not as realistic as \texttt{Sim2}, in order to better explore a wider dynamical range of parameters. To generate \texttt{Sim2}, we first follow the same method as outlined above with the following new additional steps.
First, we develop our treatment of SNR to be more realistic. Rather than setting one value of SNR for all galaxies in all bands, we use measured relationships between SNR and band magnitude in real galaxies to determine SNR for each data point. Using the data used in \cite{MassStepDust} as well as for spatially matched galaxies in the Legacy Survey data, a linear fit between magnitude and SNR was calculated for every band in this work. Then, in the model creation process, each data point for each galaxy is assigned a SNR determined using its magnitude and the derived linear relationship. An error is then drawn randomly from a Gaussian distribution centered at 0 with standard deviation of flux divided by this SNR. This gives this sample of models a distribution of SNR in each band akin to a real galaxy dataset.

In addition, we replace our discrete $A_V$ distribution with a continuous one. Instead of looping through five values of $A_V$ as we create our models, we select five random values of $A_V$ from a uniform distribution ranging from 0.3 to 1.7 for each parameter combination. To clarify, whereas in our \texttt{Sim1} sample there were five copies of every galaxy which only differed by their value of $A_V$, there are still five copies of every galaxy which only vary in their $A_V$ value but now these values are randomly selected rather than predetermined. $\delta$ is now also a continuous parameter, randomly drawn between -1.6 and 0.4. We distinguish between \texttt{Sim2}, where  both $\delta$ and $A_V$ take a range of independent values, \texttt{Sim2a}, where $\delta$ is fixed, and \texttt{Sim2b}, where $\delta$ is calculated from $A_V$ based on the \citet{salim2020dust} empirical relation described in more detail below.
After models are generated we apply a series of cuts to mimic a realistic dataset. We first select only galaxies with a sSFR (the specific star formation rate, which is SFR divided by the stellar mass) in the range $10^{-13} <$ sSFR $< 2 \times 10^{-9}$ as these are typically expected values (see, e.g., \citealt{sSFR}).
We then also enforce that the magnitude in the $i$ and $r$ bands be consistent with observed values. We cut any galaxies in our sample that do not fall between the 1st and 99th percentiles for the $i$ and $r$ bands in the \cite{MassStepDust} data. Due to the now random nature of the model creation process, a slightly different number of models passes the selection criteria for each sample, and our \texttt{Sim2} sample size ranges from 2,883 to 2,945 galaxies for the different cases outlined in \S \ref{sec:relations}.

\begin{table*}
\begin{tabular}{llll}
\hline
\hline
Parameter & Symbol / Unit & \texttt{Sim1} Input Values & Fit Prior Distributions\\  
\hline
Redshift & $z$ & 0.2, 0.4, 0.6, 0.8 & $\cal{N}$(observed redshift, 0.001) \\
Mass Formed  & log$_{10}(M_{\rm form}\ /\ \mathrm{M_\odot})$ & 8.5, 9.5, 10.5, 11.5 & ($1 - 15$) Logarithmic \\ 
Metallicity & $Z\ /\ \mathrm{Z_\odot}$ & 0.5, 1, 2 & (0, 2.5) Uniform \\
\hline
\hline
$V$-band dust attenuation & $A_V$  & 0.100, 0.525, 0.950, 1.375, 1.800 & (0, 2) Uniform \\ 
Deviation from Calzetti slope & $\delta$ & -1.400, -0.975, -0.550, -0.125, 0.300 & (-1.6, 0.4) Uniform  \\
\hline
\hline
log-normal SFH max time & $t_{\rm max}$ / Gyr & 2, 4.66, 7.33, 10 & (0.1, 15) Uniform\\
log-normal SFH FWHM & $\sigma_{\rm SFR}$ / Gyr & 4, 7.33, 10.66, 14 & (0.1, 20) Uniform \\
\hline
\hline
Strength of the 2175 \AA~bump & B & n/a & (0.15, 0.85) Uniform \\ 
\hline
\hline
Minimum starlight intensity dust receives & U$_{\mathrm{min}}$ & n/a & (0.1, 24) Logarithmic \\
Dust fraction receiving intensity $>$ U$_{\mathrm{min}}$ & $\gamma$ & n/a & (0.0005, 1) Logarithmic \\
PAH mass fraction & q$_{\mathrm{PAH}}$ & n/a & (0.1, 4.6) Logarithmic \\ 
\hline
\hline
\end{tabular}\caption{Host galaxy parameters values entered into \texttt{BAGPIPES} to simulate galaxies in the \texttt{Sim1} set (third column), along with the priors assumed in the SED fitting (last column). }\label{table:params}
\end{table*}

Moreover, we make simulations that include the 2175 \AA~ bump. We use the \texttt{Sim2} model setup but draw triple the number of galaxies with each galaxy given a randomly drawn bump strength value in the range (0.15, 0.85); this ensures that for each combination of all other parameters we have three unique bump strength values.

\texttt{BAGPIPES} allows for treatment of dust's re-emission of absorbed light as parameterized by \cite{Emission}.
Their model uses three parameters,  $q_{\mathrm{PAH}}$, $U_{\mathrm{min}}$, and $\gamma$. PAH stands for polycyclic aromatic hydrocarbon, and $q_{\mathrm{PAH}}$ is the mass fraction of the dust in the form of these molecules. $U_{\mathrm{min}}$ is the minimum of the distribution of starlight intensity that the dust is exposed to, and $\gamma$ is the mass fraction of the dust that is heated by starlight with an intensity above the minimum $U_{\mathrm{min}}$ \citep{Emission}. $U_{\mathrm{min}}$ is defined as a scaling factor of the intensity of the
interstellar radiation field as estimated by \cite{Mathis1983} for the solar neighborhood, meaning it is unitless like $\gamma$ and $q_{\mathrm{PAH}}$.

To analyze the residual distributions of these parameters, we again use the \texttt{Sim2} model setup {using all photometric bands} but draw triple the number of galaxies, where each galaxy triplet has the same combination of all other parameters, but a unique dust emission value. Note that if we were to take this approach for each dust emission parameter, we would end up with a factor of 27 more models than in the default \texttt{Sim2} case, which is computationally expensive to the point of being prohibitive. Thus, we instead draw three random sets of the emission parameters, rather than drawing three of each parameter and considering every possible combination. We assume input values ranging over the values taken from \cite{Emission} ($q_{\mathrm{PAH}}$ from 0.1 to 4.6, $U_{\mathrm{min}}$ from 0.1 to 24, $\gamma$ from 0.0005 to 1). Our final set of \texttt{Sim2} simulation includes 2 different subsets of simulations: one with no dust emission, one with dust emission.

To conclude, our landscape of \texttt{Sim2} simulations includes:  
\begin{itemize}
    \item five different band combinations
    \item three different $\delta - A_V$ relations (unconstrained, slope 0, and empirical slope)
    \item two bump schemes (no bump, bump)
    \item two dust emission schemes (no emission, variable Draine \& Li parameters).
\end{itemize}

\subsection{SED fitting}

\begin{figure*}
    \includegraphics[width=1\textwidth]{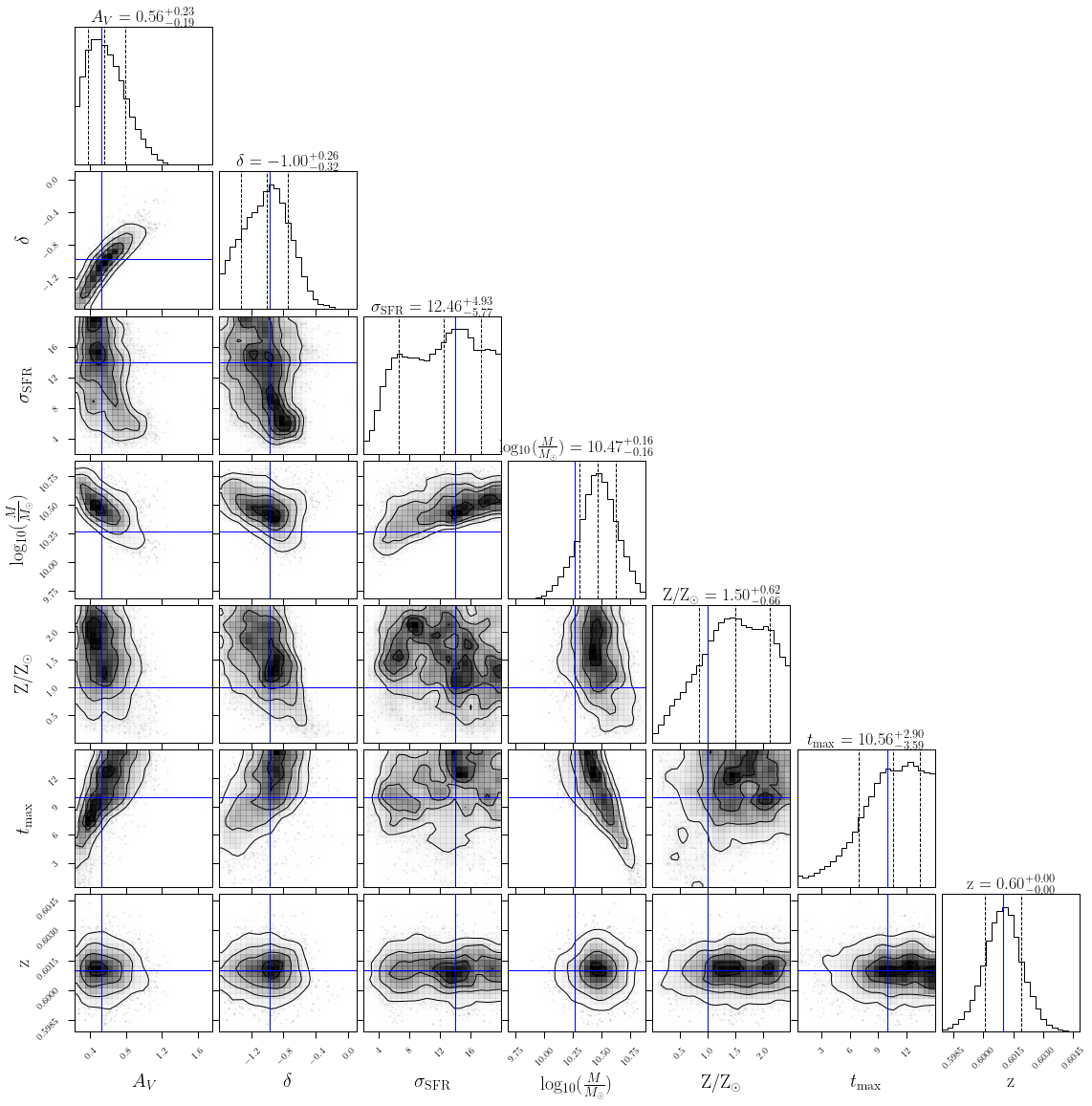}
    \caption{Corner plot of each of the 2D and 1D posteriors for all 7 parameters considered for an example galaxy, using $ugrizJHK_s$ bands. The truth value for each parameter is plotted as the blue line, while the 16th, 50th, and 84th percentile of the recovered posterior are shown in the diagonal 1D plots as dashed vertical lines. The truth values for this galaxy were $A_V = 0.525$, $\delta = -0.975$, $\sigma_{\mathrm{SFR}} = 14.0$ Gyr, $\mathrm{log}_{10}(M_{\star}/M_{\odot}) = 10.27$, $Z/Z_{\odot} = 1.0$, $t_{\mathrm{max}} = 10.0$ Gyr, $z = 0.601$. }
    \label{corner}
\end{figure*}

\begin{figure*}
    \includegraphics[width=1\textwidth]{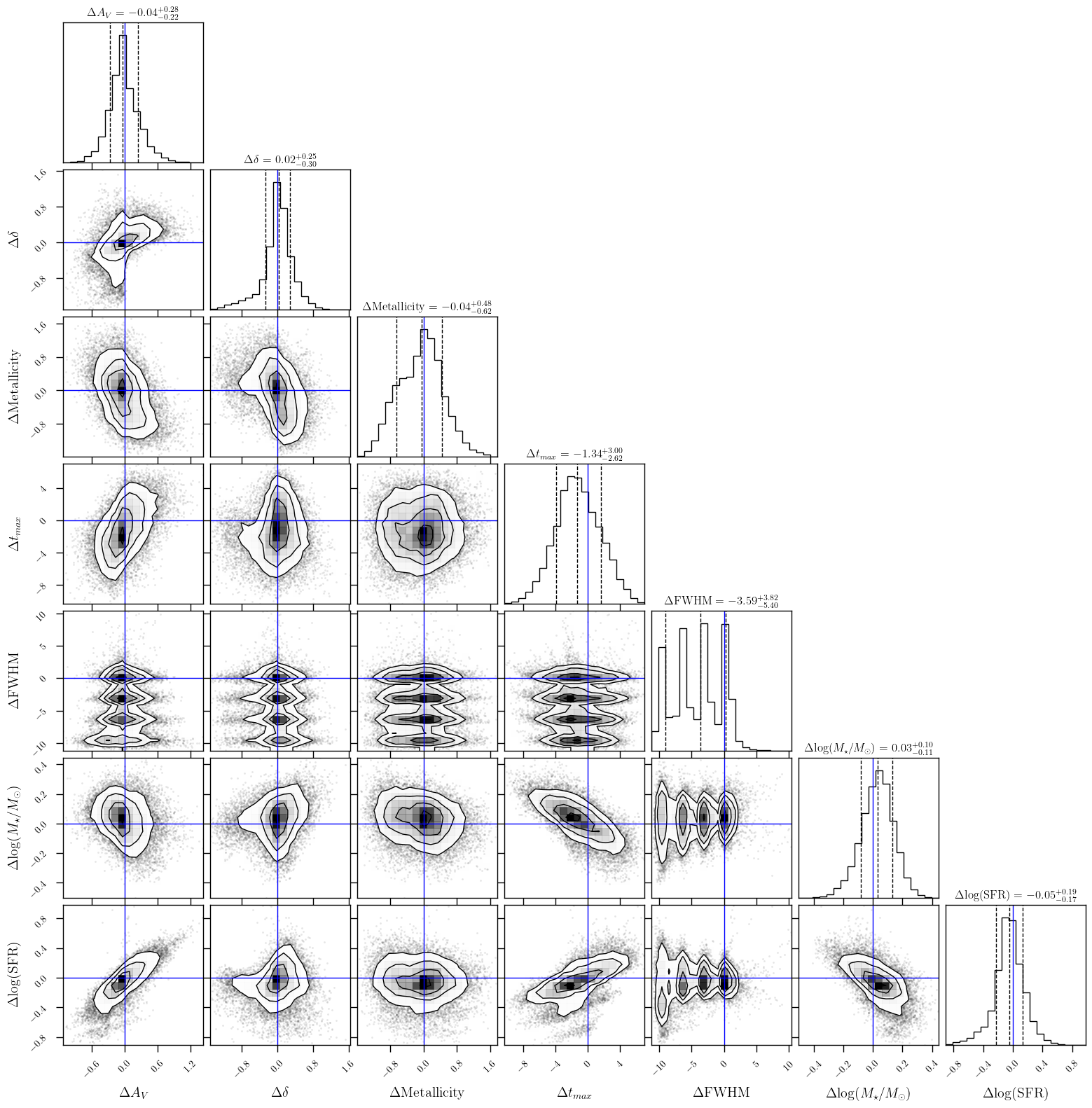}
    \caption{Corner plot of the residuals (the truth value minus the recovered value) for each of the seven parameters considered, using $ugrizJHK_s$ bands. Each solid line represents zero (i.e. an accurate recovery of the input parameter), while the dotted lines in the 1-dimensional histograms represent the median and the 32nd and 68th percentile of the distribution. In order to retain clarity, a $\Delta$log(SFR) $<1$ dex cut has been applied, removing 66 of 19,200 galaxies. }
    \label{corner_residuals}
\end{figure*}

The simulated photometric data of these galaxies are then fed into  \texttt{BAGPIPES} so that we can extract galaxy parameters as though they were real data. We assume the same Salim dust law and lognormal SFH as was used to generate the models in order to directly determine how accurately specific parameters are recovered. {Prior distributions are chosen such that the prior encapsulates the entire range of possible generated values.} Prior distributions on dust parameters are all uniform, ranging from 0 to 2 and -1.6 to 0.4 for $A_V$ and $\delta$ respectively. The SFH priors are also uniform, where the prior on $t_{\rm max}$ extends from 0.1 to 15 Gyr, and the full-width-half-maximum prior covers 0.1 to 20 Gyr. The prior on metallicity allows for values between 0 and 2.5, and is again uniform. The prior on the mass formed ranges from $10^1$ to $10^{15} M_{\odot}$ and is uniform in logarithmic space. Finally, the prior for the redshift is a gaussian with a standard deviation of 0.001 centered at the ``measured" value of redshift, i.e. the truth redshift with a simulated scatter added on. All of these priors are also given in Table \ref{table:params}. For the dust emission, we assume the prior range that follows the input for the simulation ($q_{\mathrm{PAH}}$ from 0.1 to 4.6, $U_{\mathrm{min}}$ from 0.1 to 24, $\gamma$ from 0.0005 to 1), and give each parameter a logarithmic prior in that range following \cite{Chastenet_2019}.



\section{Results}\label{sec:results}

An example of a corner plot of resultant posteriors for one galaxy is shown in Figure \ref{corner}. This galaxy was selected due to its residuals {(truth value minus recovered value)} being close to the average residual for each parameter of interest. The entered truth values for this galaxy were $A_V = 0.525$, $\delta = -0.975$, $\sigma_{\mathrm{SFR}} = 14.0$, $\mathrm{log}_{10}(M_{\star}/M_{\odot}) = 10.27$, $Z/Z_{\odot} = 1.0$, $t_{\mathrm{max}} = 10.0$, $z = 0.601$. Observing the posteriors for each parameter will give the reader an idea of the degeneracies and uncertainties present in the SED fitting process. For instance, the broad joint 2D posteriors between metallicity and $t_{\mathrm{max}}$ or $\sigma_{\mathrm{SFR}}$ reflects the difficulty in fitting galaxy age and metallicity \citep{AgeMetal}. The stellar mass posterior is strongly correlated with the SFH parameters ($t_{\mathrm{max}}$ or $\sigma_{\mathrm{SFR}}$) as the stellar mass is calculated from the SFH. Finally, one can see that there does exist a strong degeneracy between $A_V$ and $\delta$ as is to be expected. It remains to seem whether this fitting degeneracy is causing false correlations to emerge in the fitted data. A somewhat milder degeneracy is also present between stellar mass and the dust parameter. 

Note that unless otherwise specified, in the following dust emission is set to 0 and not fit for. This is because only the longest wavelength band considered here, W4, is significantly affected by dust emission and does not significantly contribute to the dust absorption measurements. In other words, including or excluding W4 in the following where we have turned off dust emission does not have an impact on our results, as it is confirmed by our findings below. For similar reasons, we initially use the simulations with no bump. We separately assess the impact of variable bump strength and dust emission parameters on our results in Sec. \ref{sec:Bump} and Appendix \ref{sec:emission}.

We now focus on the residual distributions of the parameters defined as the difference between the truth and the recovered values. The distributions of the residuals for each of the seven parameters is shown in Figure \ref{corner_residuals} for galaxies from the \texttt{Sim1} data set, using the $ugrizJHK_s$ bands. All of the residual distributions follow roughly Gaussian distributions with the exception of the SFH full width half maximum ($\sigma_{\rm SFH}$), which has four discrete peaks. This seems to be due to the fact that recovered $\sigma_{\rm SFH}$ is uncorrelated with the true $\sigma_{\rm SFH}$. For the majority of models, regardless of their $\sigma_{\rm SFH}$ truth value, \texttt{BAGPIPES} tends to yield a $\sigma_{\rm SFH}$ near a value of 13.5 Gyr, corresponding to a very flat SFH, with large uncertainties, averaging to 4.8 Gyr. This is a consequence of our inability to reconstruct the entire SFH from photometric data, which is not surprising \citep{Carnall_2019,Leja_2019}.  While only 14\% of galaxy models were given input $\sigma_{\rm SFH}$ values within 1 Gyr of 13.5, 75\% were calculated to have $\sigma_{\rm SFH}$ values in this range. This tendency for all galaxies to recover the same $\sigma_{\rm SFH}$ results in the multi-peaked distribution we recover; since the recovered value is almost constant and the truth values are discrete, the residual distribution becomes almost discrete. Because estimating this parameter is meaningless in this analysis, we ignore it in the following. We will also not focus on any SFH parameters, except for the SFR, which is the only star formation parameter that can be more reasonably estimated (with a $0.18$ dex scatter in \texttt{Sim1}) from photometric SED fitting.

For all the other parameters, it is remarkable that the bias distributions have a median which is close to 0, and in all cases within far less than the 1$\sigma$ of the distribution. Despite the degeneracies, the stellar mass bias only has a 0.1 dex scatter (here considered as half of the central 68 percentile of the bias distribution), close to a typical stellar mass uncertainty (0.11 dex on average in this sample) from this type of measurement (e.g. \citealt{palmese16,palmese18, ConroyReview}), while $A_V$ and $\delta$ have a scatter of 0.25 and 0.28, respectively. The SFR scatter is below 0.2 dex, though the SFH parameters have higher scatters: $t_{\mathrm{max}}$ having a scatter of 2.8 and $\sigma_{\mathrm{SFR}}$ having a scatter of 4.6.

Considering the 2D distributions, some parameters show correlations that require further investigation. However, certain correlations, such as the correlation between the $t_{\mathrm{max}}$ and SFR residuals, are expected due to the fact that SFR is calculated directly from $t_{\mathrm{max}}$ and $\sigma_{\mathrm{SFR}}$. Also the correlation between the $A_V$ and SFR residuals can be understood since the dust attenuation in UV corrects the observed UV luminosity to get SFR. Given the direction of the degeneracy, the correlation is due to a confusion between ``red and dead'', low SFR galaxies with low dust content, and young, high SFR, high dust content galaxies. The $A_V$-SFR correlation will be explored in the following subsection. 

We note that the SNR used in \texttt{Sim1} is worse than that from the more realistic distributions in \texttt{Sim2}, therefore also the recovered parameters will have more scatter. By measuring the parameters residuals with \texttt{Sim2} we recover scatters that are half or less than those we present for \texttt{Sim1}. A similar problem persists for the SFH FWHM.

\subsection{Dust parameters and star formation rate}

\begin{figure}
    \includegraphics[width=0.4\textwidth]{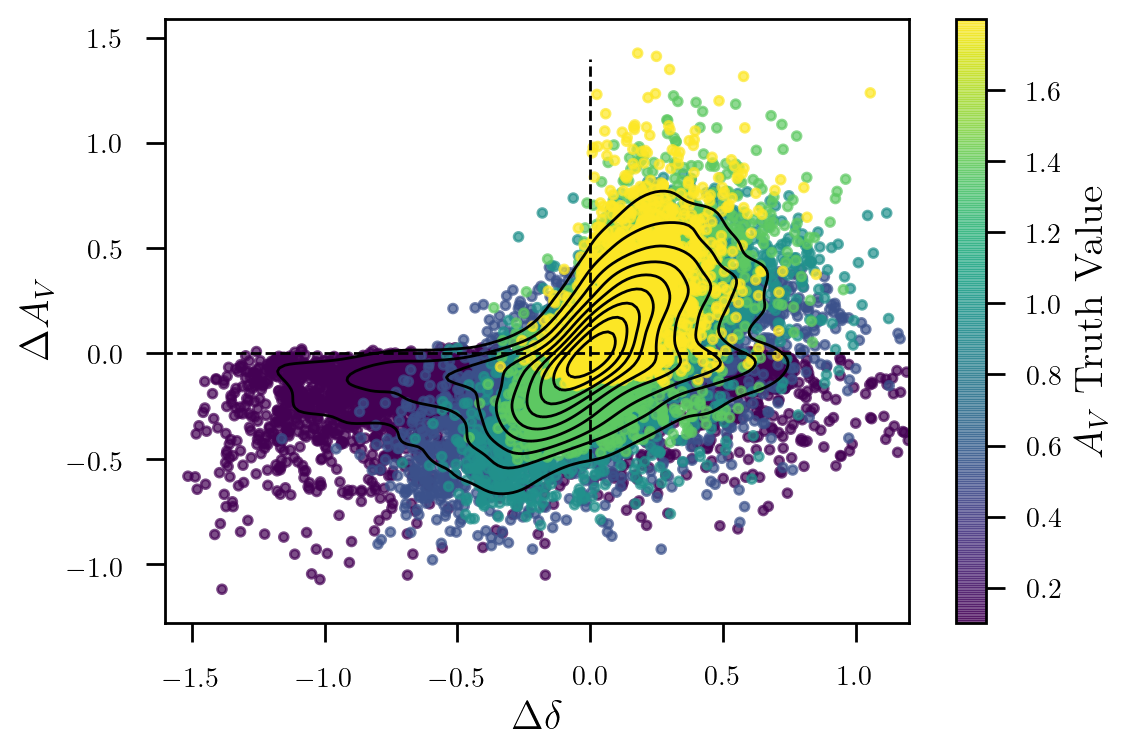}

    \caption{Recovered $A_V$ and $\delta$ residuals (the truth value minus the recovered value) using $ugrizJHK_s$ bands. The truth value for $A_V$ is given by the color bar, while the black dashed lines indicate zero residual, i.e. a perfect recovery of the truth value. The black solid lines indicate the percentiles of the joint distribution from 10 to 90\% in steps of 10\%.}
    \label{AvDeltaContour}
\end{figure}

\begin{figure*}
    \includegraphics[width=1\textwidth]{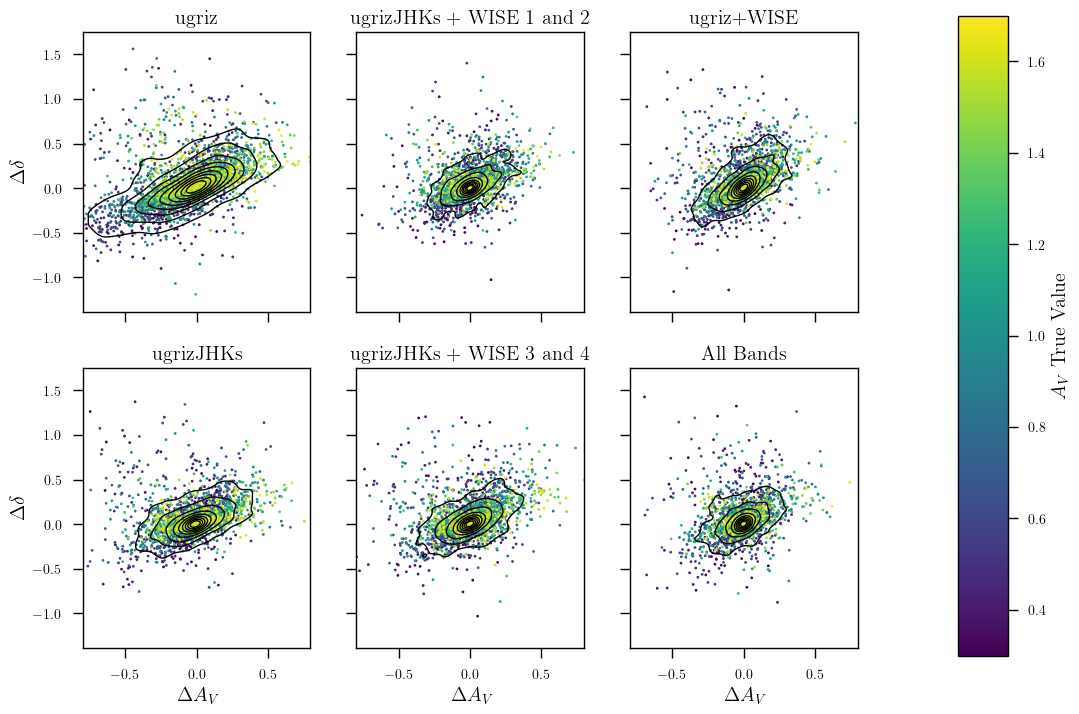}
    \caption{Recovered $A_V$ and $\delta$ residuals (the truth value minus the recovered value) using \texttt{Sim2} galaxies in six combinations of bands. The truth value for $A_V$ is given by the color bar and the black solid lines indicate the percentiles of the joint distribution from 10 to 90\% in steps on 10\%.}
    \label{AvDeltaContourAllBands}
\end{figure*}

In Figure \ref{AvDeltaContour}, we show the recovered distribution of $A_V$ and $\delta$ residuals for our model sample \texttt{Sim1}. The contour lines represent the percentiles of the joint distribution from 10 to 90\%, and reveal a correlation between $\Delta A_V$ and $\Delta \delta$, meaning that an overestimation in $A_V$ is likely to be paired with an overestimation in $\delta$, and the same with an underestimation. A degeneracy is expected due to the definition of these parameters. Because the attenuation is always better estimated when comparing shorter wavelengths to the rest frame NIR, where the effect of dust is minimal, is it reasonable to expect that an overestimation of $A_V$ will also lead to an overestimation of $\delta$, while the NIR data points provide a ``zero-point' to the estimated attenuation law. One can also understand this in the following way: the total dust luminosity (the difference between the dust-free SED and the observed SED) is better constrained than either $A_V$ or $\delta$ individually. In order to keep the total dust luminosity constant, the slope of the attenuation law must decrease, which amounts to making $\delta$ larger. Another relevant aspect to note to understand why we cannot recover the dust parameters very precisely is that at the redshifts considered here, with $u$ being the shortest band, we are not sampling the rest-frame far UV. The lack of FUV will degrade dust estimation (especially since IR covering the typical dust emission wavelengths is not available either).

The fact that the contour lines elongate along the x-axis close to $\Delta A_V = 0$ are due to regions where the input value of $A_V \simeq 0$, as the colors in Fig. \ref{AvDeltaContour} show. A low $A_V$ corresponds to cases with little dust (or where the effect of dust on the SED is less pronounced), so that $\delta$ is difficult to constrain, and it appears unconstrained from our fits. Because low $A_V$ values (truth $A_V=0.1$) lie at one of the edges of the prior (which only allows positive values down to 0), it is expected that most low $A_V$ galaxies will lie on the $\Delta A_V < 0$ side rather than above 0. 

We next seek to explore how the dust residual degeneracies change with the amount of data used. Now moving to our \texttt{Sim2} sample of galaxies to represent a more realistic galaxy population and considering different combinations of bands used, we generate Figure \ref{AvDeltaContourAllBands}. The effect of more data is immediately evident, the inclusion of more bands (to clarify, we are comparing less versus more bands both using \texttt{Sim2}, these changes are not due to changing from \texttt{Sim1} to \texttt{Sim2}) both reduces the size of the scatter in $\Delta A_V$ and $\Delta \delta$ (from 0.25 and 0.26 in the $ugriz$ case to 0.12 and 0.17 respectively in the All Bands case) while also lifting the degeneracy between the two parameters, as can be determined by comparing the shapes of the contour lines. We find that the inclusion of W1 and W2 provides a slightly better result for the scatter (0.12 for $A_V$ and 0.15 for $\delta$) than the addition of W3 and W3 (0.14 for $A_V$ and 0.17 for $\delta$) to $ugrizJHKs$ for both $A_V$ and $\delta$, while the bias is similar and consistent with 0 for the two cases. In addition, the correlation between the true $A_V$ value and the $A_V$ residual is still present in these plots, though it is not as extreme in the Figure \ref{AvDeltaContour} case, likely due to fewer low $A_V$ galaxies being present in the \texttt{Sim2} case (as the distribution of values peaks above 0.1), amongst other effects. The correlation between the true $A_V$ value and the $A_V$ residual also lessens as more bands are added. This is likely due to the fact that increasing bands allows us to measure both $\delta$ and $A_V$ more precisely, reducing the error compared to the range of values they can take and hence ensuring against an artificial trend.

\begin{figure*}
    \includegraphics[width=0.7\textwidth]{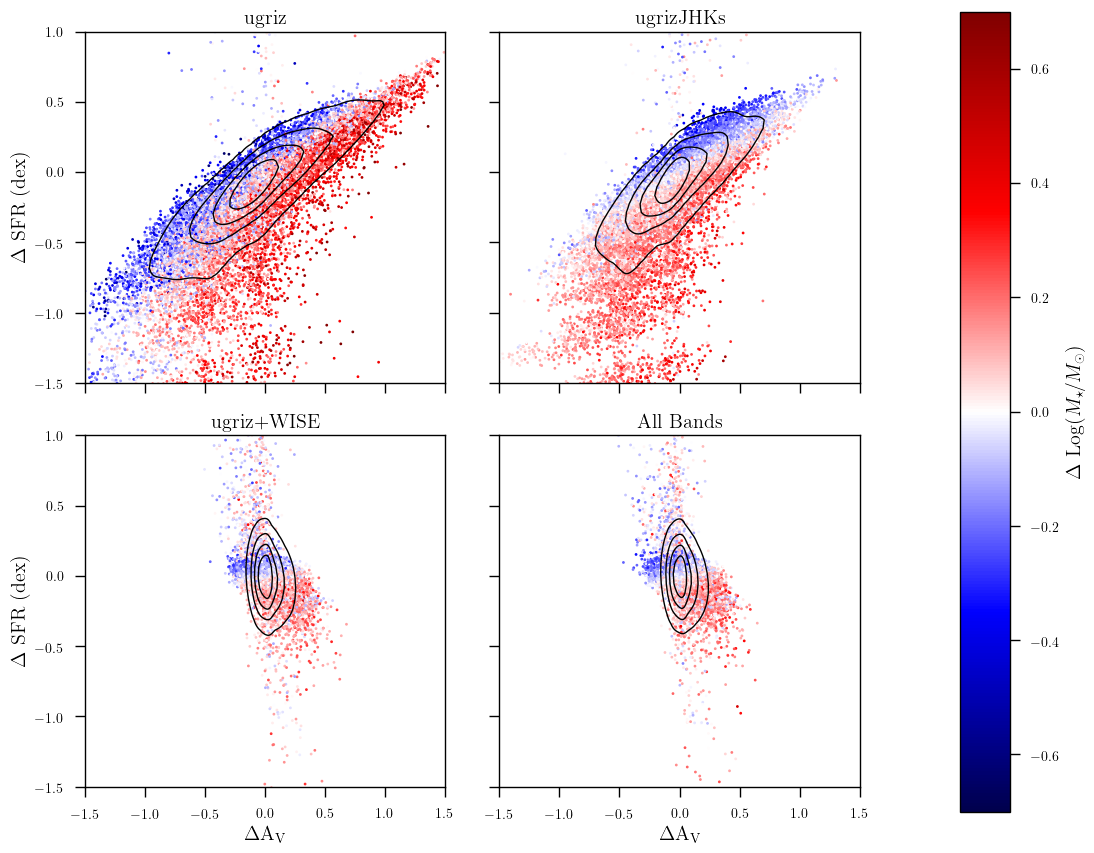}
    
    \caption{Residuals of $A_V$ versus residuals of SFR, where each subplot represents a different choice of photometric bands. The $20$, $40$, $60$, and $80$ percent contour lines are plotted in black. The degeneracy between $A_V$ and SFR residuals is evident in the upper two plots, but the degeneracy is reduced by the use of additional bands. The stellar mass residual is also plotted via the color of the points, demonstrating that any $A_V$ - stellar mass degeneracy is also lifted by the use of additional bands. SFR and stellar mass are correlated in all plots, which is not surprising as stellar mass is dependent on the SFR by definition.}
    \label{AvSFR}
\end{figure*}

Other degeneracies in residuals exist, notably a degeneracy in the residual of $A_V$ and of SFR. In addition, both of these residuals seem to correlate with stellar mass as well. To investigate this relationship, we compare the degeneracies with four different photometric band sets: the standard $ugriz$, $ugriz$ with the near-infrared $JHK_s$, $ugrizJHK_s$ with the WISE bands, and $ugriz$ with WISE bands. Plotting these three parameters together in the four different cases yields Figure \ref{AvSFR}. In each panel, we show the SFR residual versus the $A_V$ residual for all the galaxies in \texttt{Sim1}, while the stellar mass residual is given by the color.

The most immediately evident conclusion from Figure \ref{AvSFR} is that a degeneracy between $A_V$ and SFR residuals exists, but it can be significantly reduced by introducing additional photometric bands. It is clear that the driving factor in reducing the degeneracy is the WISE bands. Comparing $ugriz$ versus $ugrizJHK_s$ and $ugriz$+WISE versus All Bands subplots we can see there is hardly any difference between these 2 sets of plots, but the difference driven by the introduction of the WISE bands is drastic. The $ugriz$ case results in a $\Delta A_V$ scatter of $0.35$ and using $ugrizJHK_s$ gives a scatter of $0.29$, while including WISE bands reduces scatter by ~80\%, $ugriz$+WISE gives a scatter of $0.062$, the inclusion of all bands reduces it slightly further to $0.059$. This means an increase in SFR compared to the truth value can produce a galaxy SED which is similar to the true one if coupled with a larger effect of dust attenuation, namely a larger $A_V$, and vice versa with decrements in SFR and $A_V$. The addition of the two 3-5 $\mu$m WISE channels significantly helps pinning down the attenuation law further in the infrared, where dust absorption is minimal, while the inclusion of the longer wavelengths at $>10 \mu$m W3 and W4 helps to constrain the dust content via its emission. 

For what concerns the stellar mass, a degeneracy with SFR was already clear in Figure \ref{corner_residuals}, and obvious since the stellar mass is an integral of SFR over time, while a correlation with $A_V$ is less obvious. A slight degeneracy is present as photometry from a more massive, older, redder, and dust-free galaxy can be confused with a less massive, younger, dusty galaxy. Once the dust properties are more precisely constrained (via the introduction of additional bands), this degeneracy also appears to be broken, because the dust correction needed to convert the UV luminosity to SFR becomes more precise.
Simultaneously, the stellar mass scatter is significantly reduced. In the $ugriz$ bands case, the distribution has a median and 68th percentiles of $0.03^{+0.20}_{-0.19}$, which is reduced to $0.001^{+0.08}_{-0.08}$ when all bands are included. Scatter and bias are similar whether $JHKs$ or WISE bands are added to $ugriz$, so it does not seem that either of these 2 sets of bands is really driving the improvement in stellar mass.

\begin{figure*}
    \includegraphics[width=0.47\textwidth]{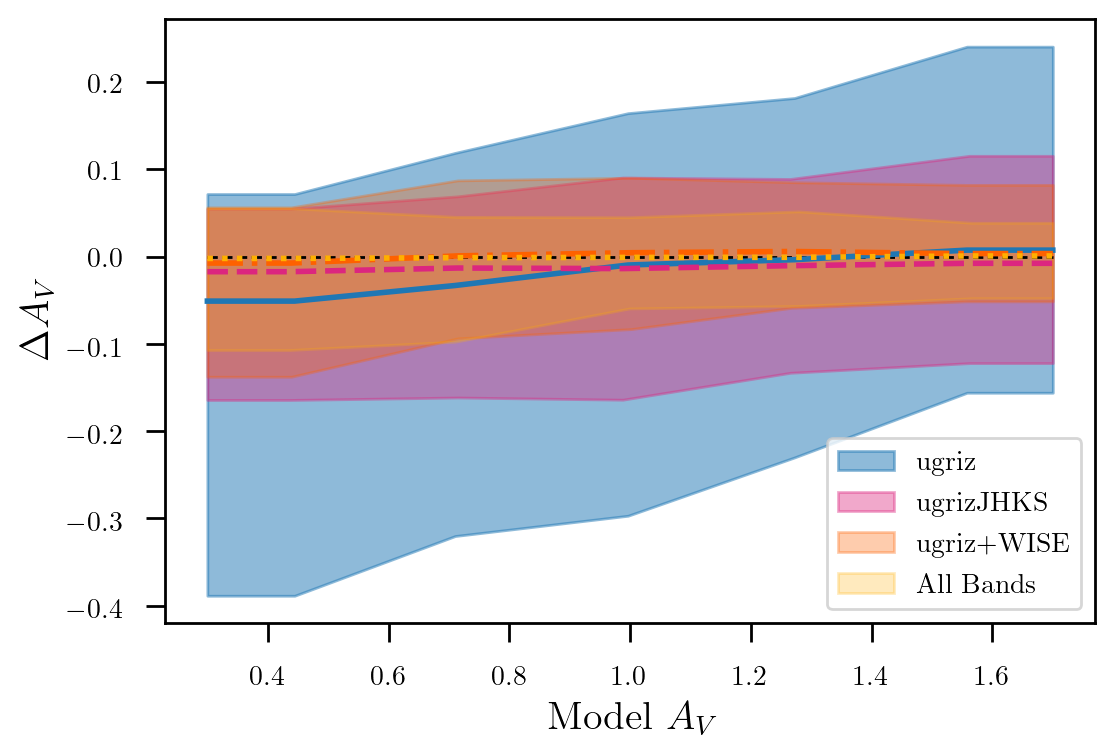}
    \includegraphics[width=0.47\textwidth]{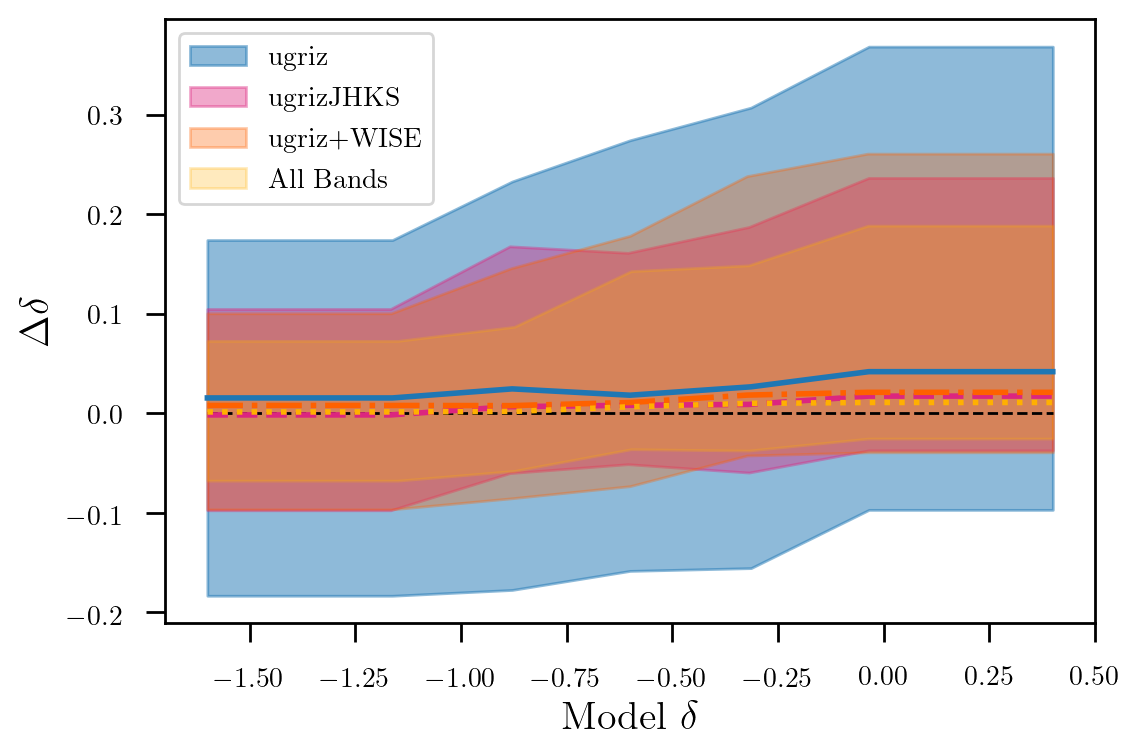}
    \caption{$A_V$ and $\delta$ residuals versus their truth value for the four different band combinations considered in this work. The central dotted lines give the median value, while the shaded region is the $1\sigma$ confidence interval. Inclusion of additional bands narrows the distribution of residuals around zero, indicating more accurate estimates. In addition, the prior tends to cause the distribution of residuals to be non-symmetric about zero and shift with increasing truth value; the inclusion of more bands significantly alleviates this effect. Note that for bins on the edge of the plot, we extend the value taken at the center of the bin to the edges of the plot to represent the entire range of dust values considered; this means that the data does not spontaneously flatten out as the plot seems to imply. }
    \label{Avcolorplot}
\end{figure*}

We next consider the effect that addition of bands has on the general recovery of both $A_V$ and $\delta$. In Figure \ref{Avcolorplot}, we plot the difference between truth and recovered values for $A_V$ and $\delta$ against their truth values. We bin the $A_V$ and $\delta$ range into five bins, and plot the median $\Delta$parameter value in each bin as lines. The shaded regions represent the 16th to 84th percentile in each bin. From this figure, we can immediately discern several positive effects that the inclusion of extra bands has on recovered values. In both the $A_V$ and $\delta$ plots the $ugriz$ relationship exhibits some degree of correlation between the truth value and the residuals. We can see that in both cases the addition of more bands reduces this correlation, nearly eliminating it entirely in the $A_V$ case. We find that the Pearson coefficient goes from 0.25 in the $ugriz$ case to 0.08 when all bands are used. For $\delta$, the Pearson coefficient is close to 0.18 in all cases. For both cases we note that the change of the median or 68th percentile of the distribution in the residuals over the entire $A_V$ or $\delta$ range consiered is typically of the order of 0.1 or less, and significantly smaller than both the distribution scatter and the dynamical range of the model values. With optical bands alone the dust parameters posteriors are significantly prior dominated, and the effect of the prior edges show up as correlations in this parameter space. These results show that analyses attempting to recover dust attenuation parameters from these optical bands alone (e.g. \citealt{duarte2022msf}) should be taken with caution. The inclusion of more bands tightens the distribution in residuals around the zero value. Each successive addition of more bands reduces the width of the 16th-84th percentile regions. Thus, with the addition of more bands not only is the median residual value closer to zero, the scatter, as expected, also shrinks. 
The persistence of the $\delta$ - $\Delta \delta$ correlation even when using all of the bands available, although to a lesser extent, (see Figure \ref{Avcolorplot}) reflects the difficulty of fitting $\delta$, resulting broad posteriors for each galaxy. Thus, the trend seen in the $\delta$ plot is more likely a demonstration of the difficulty of fitting $\delta$ in general, rather than an indication of an inherent bias introduced during fitting. Nevertheless, scatter and bias are significantly improved by the addition of bands.

\begin{figure}
    \includegraphics[width=0.47\textwidth]{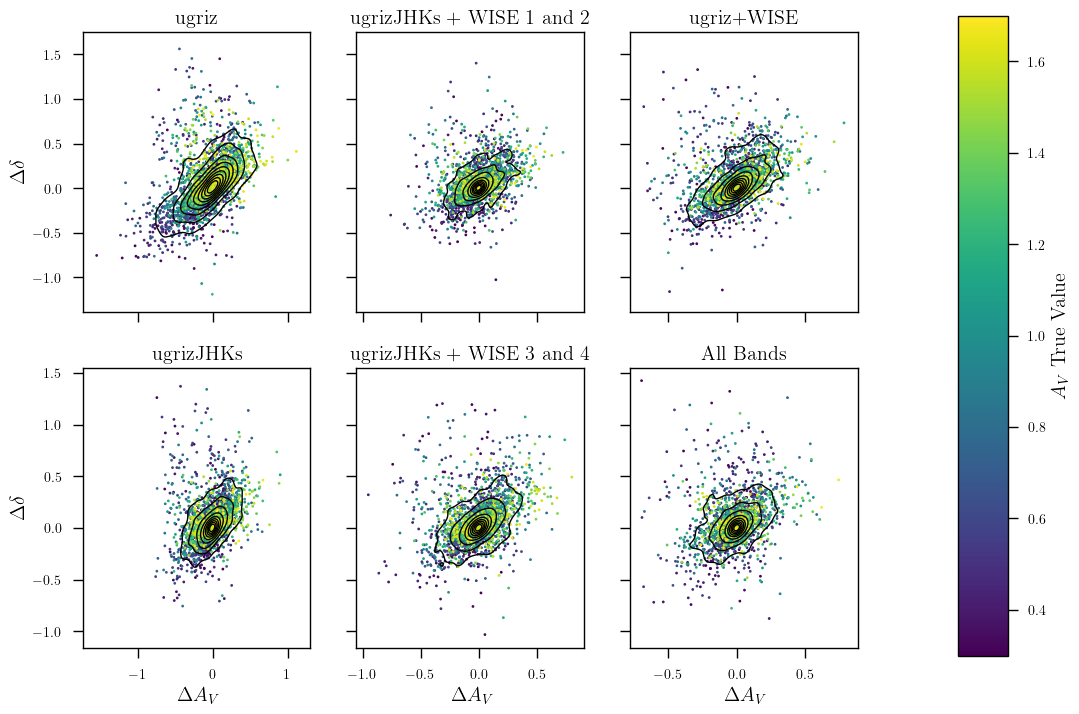}
    \caption{The median (point) and 16th and 84th percentiles (error bars) of the distribution of five galaxy parameter residuals (the truth value minus the recovered value), for four different choices of bands. Note how the inclusion of more bands both drives the median of the distribution towards zero and narrows the entire distribution.}
    \label{AllParams}
\end{figure}

Finally, we can consider the recovery of all parameters of  interest and how this is affected by choice of bands. In Figure \ref{AllParams} we plot the overall distribution of residuals for each choice of bands. Again, we notice that in general the distance from zero and the overall spread of the distribution decreases monotonically with the inclusion of additional bands. 
Considering that $ugriz+JHKs$ and $ugriz$+WISE have roughly the same number of bands, it is interesting that the latter outperforms the former in 3 out of the 5 parameters considered, while results are comparable for the other two parameters. This implies that the infrared region that WISE covers is particularly important for recovering $A_V$ and SFR, while it is less relevant for $\delta$ and stellar mass. This connection is unsurprising considering the correlation noted in Figure \ref{AvSFR}. 

\subsection{Recovering relations between $A_V$ and $\delta$}\label{sec:relations}

Recognizing and constraining relations between dust parameters can help us understand the evolution of galaxies and the physics behind dust formation. However, since dust parameters are typically extracted using SED fitting codes rather than being a direct observable from large galaxy surveys, it is a concern that any correlation measured is due to some intrinsic bias in the methodology rather than a real physical phenomenon. \cite{Qin_2022} claim that $A_V - \delta$ correlations are driven by such biases, and test their hypothesis by inputting a flat $A_V - \delta$ and demonstrating that they recover a non-flat correlation. Here, we attempt to recreate their results.

As we are now specifically checking a result that could affect measurements from real data, rather than exploring how biases in one parameter arise in conjunction with other parameters, we decide to make our galaxy sample a closer representation of realistic galaxies, and use the \texttt{Sim2} dataset. 

\begin{figure}
    \includegraphics[width=0.4\textwidth]{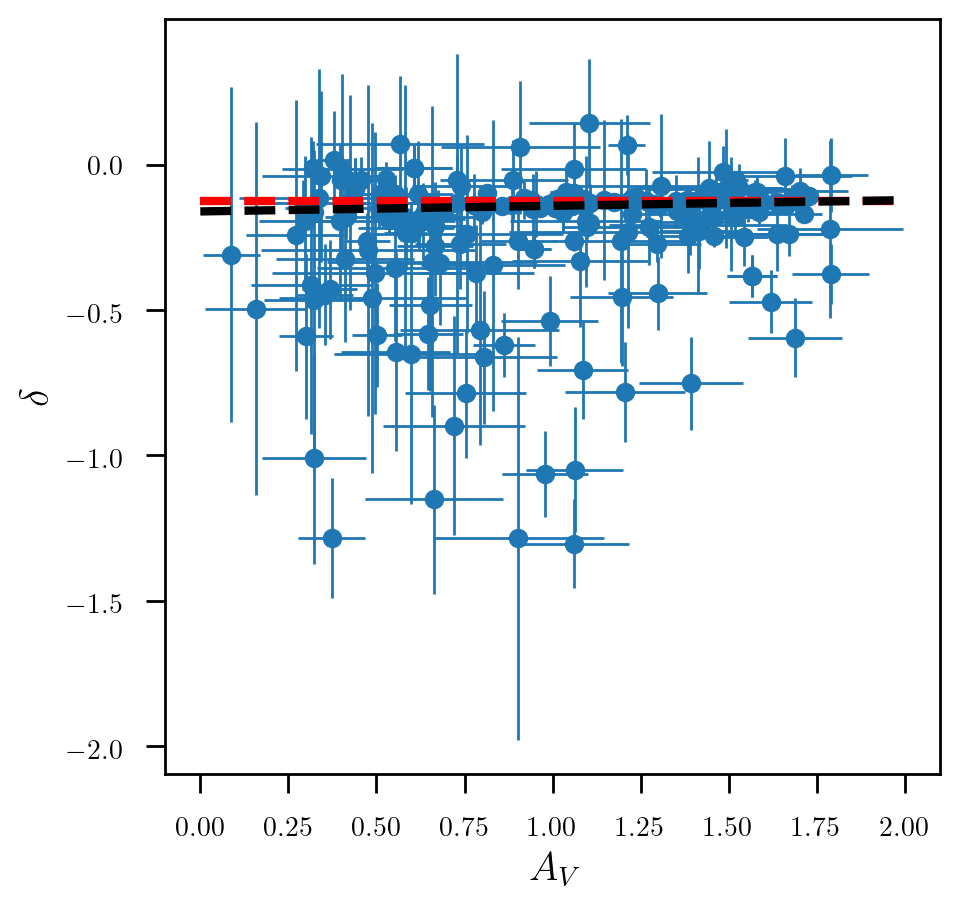}
    \includegraphics[width=0.4\textwidth]{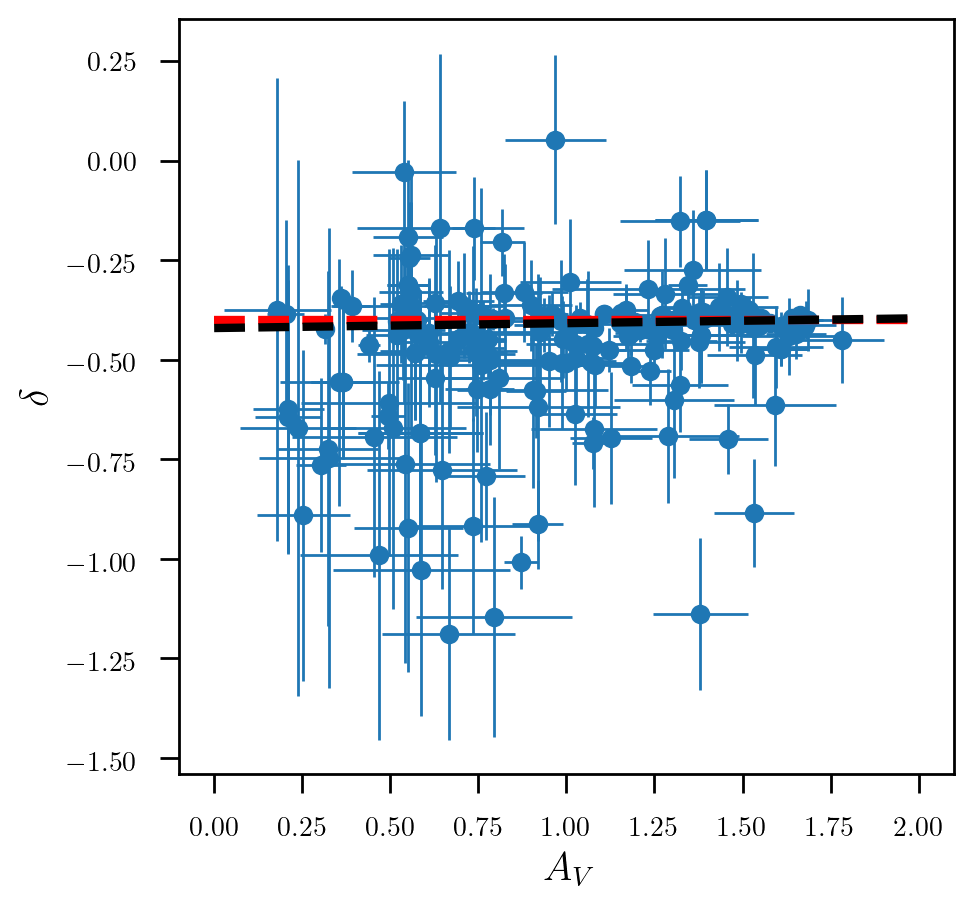}
    \caption{Results of inputting a flat $A_V - \delta$ relationship into \texttt{BAGPIPES} using all of the bands in this work and fitting a line to the output. In the top panel, we consider an input of constant $\delta = -0.125$ and in the bottom panel $\delta = -0.4$, represented by red lines. In both cases, the recovered slope (black line) is statistically consistent with zero. }
    \label{recoveredslope1}
\end{figure}

For our first experiment, we enter a flat $A_V - \delta$ relation (what we call \texttt{Sim2a}) with delta fixed at $-0.125$, similar to the value of $\delta$ used by \cite{Qin_2022}: $-0.2$. The models are calculated in the same way as the \texttt{Sim2} dataset but we fix delta to be $-0.125$ rather than selecting multiple random values. We then consider the recovered values of $A_V$ and $\delta$ after running BAGPIPES using all bands. Using a Markov Chain Monte-Carlo (MCMC) algorithm, we fit a line to the recovered data. Since the number of data points proved too large for the algorithm to compute in a reasonable amount of time, we took a random subset of two hundred galaxies. Our results are shown in the top panel of Figure \ref{recoveredslope1}.  

Here we recover a slope of m = $0.016_{-0.006}^{+0.007}$. This is consistent with zero within $2.5 \sigma$. We note that the scatter in $\delta$ is larger at low $A_V$, likely due to the fact that lower $A_V$ means that $\delta$ is harder to constrain. Since the prior on $\delta$ is set from -1.6 to 0.4, and the true $\delta$ is set to -0.125, a prior-dominated delta measurement is much more likely to have a median below its truth value due to the edge of the prior. 
To test this, we repeat the analysis with a lower $\delta$ value, -0.4. The results of this attempt are shown in the bottom panel of Figure \ref{recoveredslope1}.

In this case, the measured slope is m = 0.010$_{-0.006}^{+0.006}$. The slope is now 37.5$\%$ lower, and consistent with zero within $1.7 \sigma$. Hence, while both slopes are consistent with a flat relation, it seems that prior effects are driving only minor difference in the recovered slope. 
In \cite{Qin_2022} the recovered slope for the experiment similar to this recovers a larger slope than our results. It is unclear what is driving this difference in results. Possible explanations are different priors, the different SED-fitting codes (CIGALE versus BAGPIPES) or the different SFH models (their exponential declining versus our lognormal) used. However, the drastically improved results in our method speaks well to its accuracy in recovering correct parameter values.

\begin{figure}
    \includegraphics[width=0.4\textwidth]{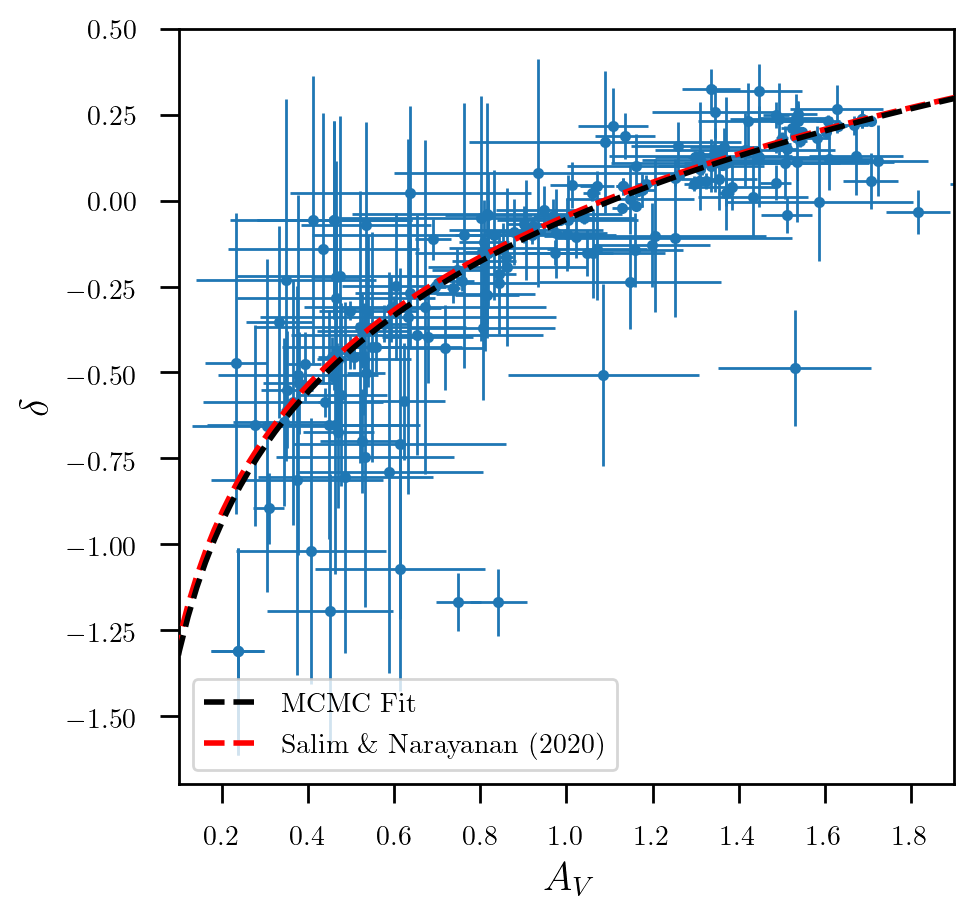}
    \caption{A similar analysis to that of Figure \ref{recoveredslope1}, though the input relationship (red line) is that derived in \citet{salim2020dust}, again using all bands this work considers. The fitted curve (black line) is statistically consistent with their results in terms of the slope of the logarithmic curve, but the intercept is shifted by a mildly significant amount.}\label{SalimPlot}
\end{figure}

Beyond simply not introducing spurious correlations, we can demonstrate that we can recover other dust parameter relations using our \texttt{BAGPIPES} runs. 
\cite{salim2020dust} derives the following relation from galaxies in their sample:
\begin{equation}
    \delta = 1.236 \times \mathrm{log}_{10}(A_V) - 0.044,
\end{equation}
which is the combination of equations (14) and (15) in that work. To determine if we can recover this distribution, we enter this relationship as truth values for another sample of galaxies (what we call \texttt{Sim2b}) and ran BAGPIPES using all of our considered bands. Our recovered fit and the expected fit is shown in Figure \ref{SalimPlot}. 
Defining m as the coefficient in front of log$(A_V)$ and b as the constant at the end of the equation, we get that m = 1.26$^{+0.02}_{-0.02}$, which varies from the \cite{salim2020dust} by only $1.2 \sigma$. Simultaneously we measure that b = -0.054$^{+0.003}_{-0.003}$. This is a statistically significant discrepancy, being slightly higher than $3 \sigma$, however it is one that is again likely driven by prior effects. As the majority of the points in this distribution are close to the upper $\delta$ prior cutoff or the lower $A_V$ cut off, this has the effect of pushing the distribution down and to the left as described above. This explains why the MCMC algorithm returns a curve shifted down from the expected curve. However, this is only introduces an error of 0.01 which is negligible when considered against the typical galaxis uncertainties in $A_V$, which are on average an order of magnitude larger.

\begin{figure}
    \includegraphics[width=0.47\textwidth]{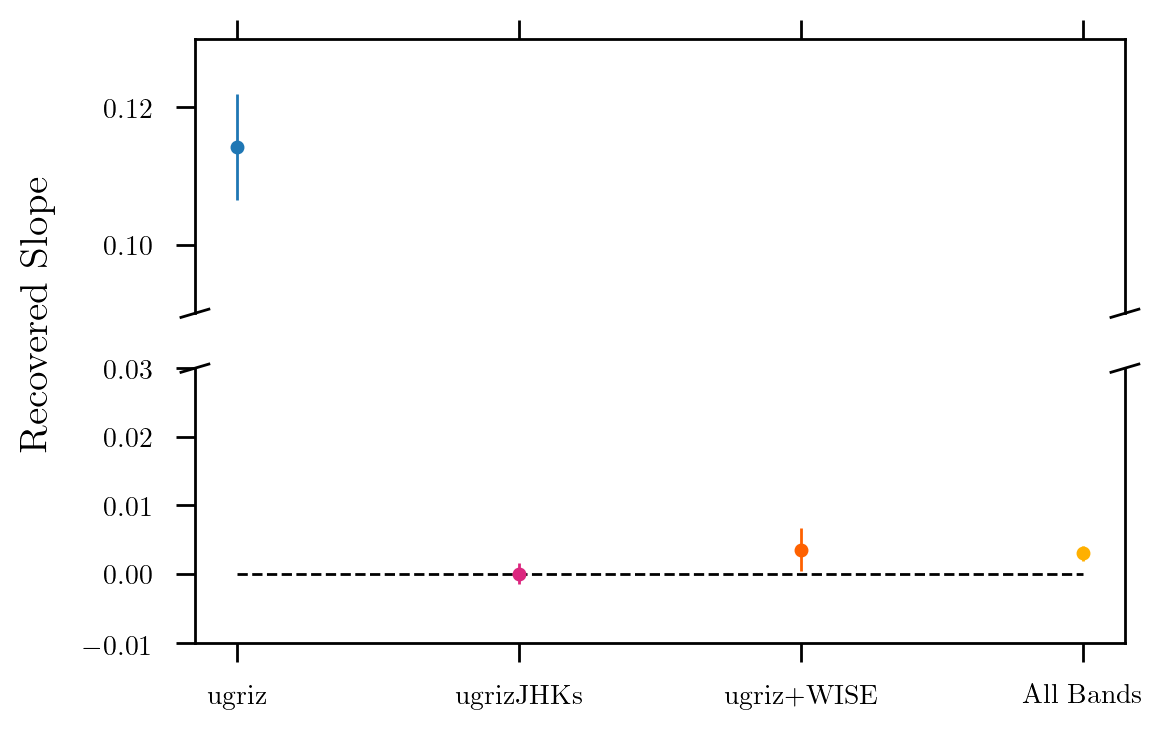}
    \caption{The recovered slope when inputting a flat $A_V$ - $\delta$ relationship at $\delta = -0.4$ for four different band combinations. With the exception of $ugriz$, all of these results are statistically consistent with zero and thus no spurious correlation is being introduced. }
    \label{slopebands}
\end{figure}

One can also consider the effect that choice of bands has on recovery of $A_V$ - $\delta$ distributions. Repeating the above flat $A_V$ - $\delta$ experiment for the four combinations of bands we have been using thus far, we generate Figure \ref{slopebands}.
Note that the axis has been broken, as the recovered slope for $ugriz$ is about 30 times larger than any other band combination. With the exception of 
$ugriz$, none of these deviations from zero are statistically significant as they are all below 3$\sigma$.

\subsection{The 2175 $\mathrm{\AA}$ bump }
\label{sec:Bump}
\begin{figure}
    \includegraphics[width=0.47\textwidth]{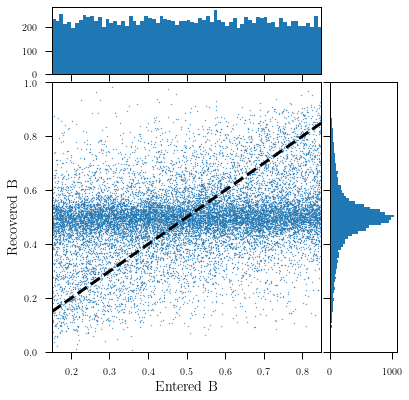}
    \caption{Distribution of entered versus recovered bump values. Though a uniform distribution of bump values is entered, the recovered bump values tend to occur in the center of the prior, indicating that the bump is extremely poorly constrained.}
    \label{BumpRecov}
\end{figure}

As mentioned above, the additional absorption in the attenuation curve known as the 2175 $\mathrm{\AA}$ bump is not considered for the majority of our analysis, as we have found previously that it had little effect on recovered parameters. We seek to reproduce and quantify that claim with model galaxies in this section. 
Immediately evident is the difficulty in recovering the value of the bump. Even when using all of the bands considered in this work, we are unable to reliably recover the value of $B$, instead returning a distribution peaked at the center of the prior (as expected since the posterior will be the same as prior and we are plotting the median); see Figure \ref{BumpRecov}. This behaviour is expected, since to better constrain the bump we would need more bands or spectroscopy at the location of the bump, rather than at longer wavelengths as we are testing here.

\begin{figure}
    \includegraphics[width=0.47\textwidth]{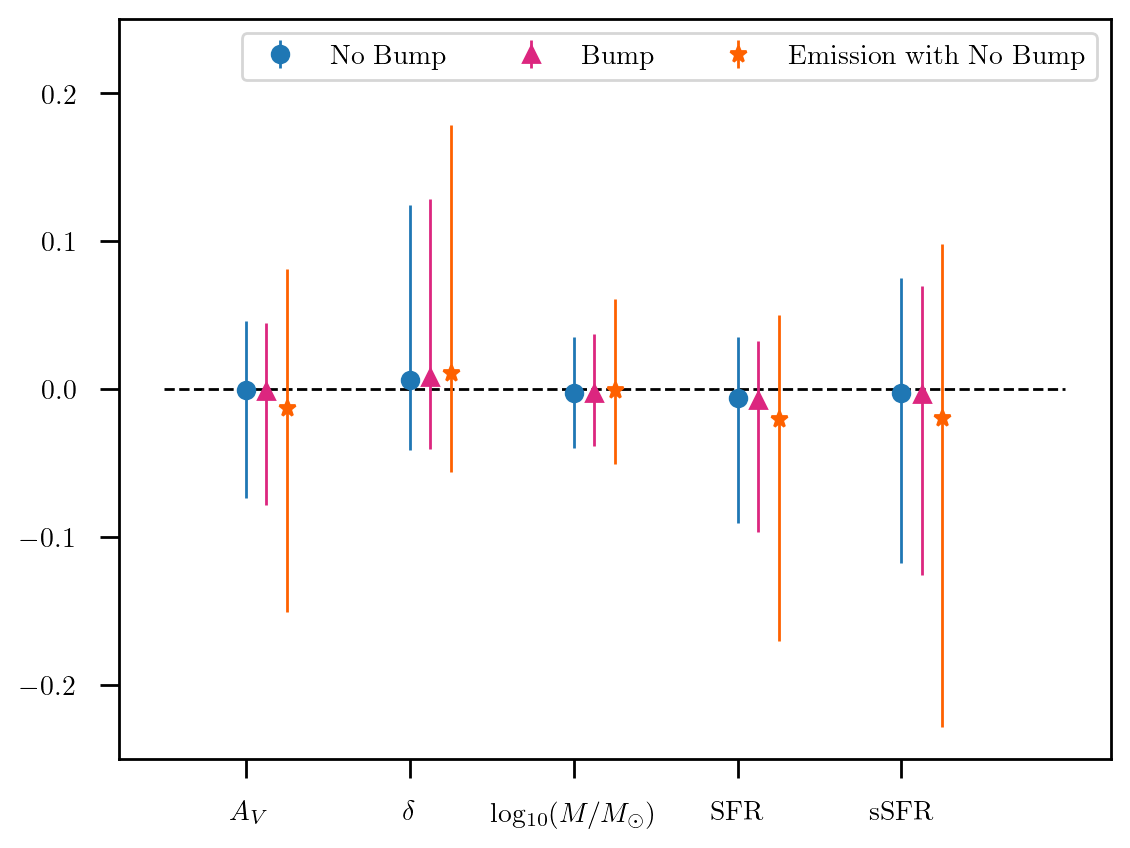}
    \caption{Residuals for the parameters of interest for different attenuation curve parameterizations, both without and with bump, as well as including IR dust emission modeling without the bump.} While the medians of all of these distributions are statistically consistent, the inclusion of emission parameters serves to increase scatter in the residuals. For example, the scatter in SFR is $0.038$ for both the default and bump cases, while it increases to $0.055$ when emission is considered.
    \label{DustTypes}
\end{figure}

Though the bump is poorly constrained itself, it does not introduce any additional error into the other parameters considered. Considering Figure \ref{DustTypes}, one can determine that the residual distributions for parameters of interest are statistically consistent when fit with and without a bump. Hence, while the bump seems to be poorly fit, including it in a SED fitting process should not harm the other parameters considered. Although the effect is small, we still suggest including the bump modeling (even if UV is not covered) because allowing for that extra attenuation does change the total dust luminosity that needs to be matched to IR luminosity for energy balance. 

\section{Conclusions}\label{sec:conclusions}
In this work, we have presented an analysis of the spectral energy distributions of a range of model galaxies using the \texttt{BAGPIPES} software. Our analysis of galaxy parameter residuals after simulating model galaxies and fitting their SEDs revealed a degeneracy between $A_V$ and SFR (as is to be expected, $A_V$ is a dust correction to SFR), which can be decreased with the inclusion of additional photometric bands.
Specifically, one can reduce $\Delta A_V$  from -0.10$^{+0.38}_{-0.40}$ to 0.01$^{+0.11}_{-0.08}$ and $\Delta$SFR from -0.12$^{+0.26}_{-0.43}$ to -0.01$^{+0.06}_{-0.06}$ by adding dust IR constraints to a case with neither rest-frame UV nor dust IR.
This trend generally extends to all parameters of interest. Comparing recovered residuals for $A_V$, $\delta$, stellar mass, SFR, and sSFR revealed the reduction in both mean residual and scatter for these parameters. 
 The inclusion of more bands reduces the correlation between entered $A_V$ or $\delta$ and their residual. This means that the addition of more bands reduces prior effects that can introduce spurious relations in one's results.
We next demonstrated that BAGPIPES does not introduce systematic biases when fitting for dust parameters. In our tests using a flat $\delta - A_V$ relation and using all bands, we always measured any deviation from expected values as being less than $3 \sigma$. Even in the scenario where only optical bands are used, any minimal correlations that we find are never as steep as the observed ones between $A_V$ and $\delta$  \citep{salim2020dust}, making it hard to reproduce the observed correlations with spurious fitting problems alone.
Similarly, BAGPIPES is able to recover physical dust parameters with distributions following those found in previous works. Entering a measured $A_V - \delta$ relationship from \cite{salim2020dust} as our truth values, we measured a recovered curve statistically consistent with the expected distribution, though the intercept is mildly biased. This discrepancy is interpreted as due to prior effects and is much smaller than typical statistical uncertainties on $\delta$.

Finally, we demonstrated that including an additional degree of freedom, in the form of a dust bump, should not drastically affect one's determination of relationships between various galaxy parameters, as its inclusion does not introduce any additional degeneracies into the fits.

Our findings are indicative of the fact that correlations such as those found in \citet{MassStepDust} between the SN Hubble residuals and dust parameters are unlikely to arise from  \texttt{BAGPIPES} fitting problems or spurious relations between parameters due to degeneracies, as we do not find evidence for large biases or spurious relations. Similar results have been found in works such as \cite{Boquien2022}.

It is important to note that some simplifying assumptions have been made in this work. In particular, we have produced and fit the simulated data with the same models in most of the analyses: in reality this may not be the case. In other words, we do not know if, in general, the real attenuation law or SFH are well approximated by a Salim law with no bump and a lognormal parametrization, respectively. On the other hand, it is worth noting that our findings show that the SFH parameterization is decoupled from the dust constraints, and even when SFH parameters are barely constrained, we can recover dust parameters within 0.2-0.3 dex. Moreover, our tests in which we assume a 2175 \AA~ bump and  then fit using the models that do not have it, still show we derive a $A_V, \delta$ accuracy  similar to the case with no bump,
meaning that at least for what concerns the bump, a simplified dust model does not need to capture all the features of the dust law to recover its slope and normalization.

We have not explored the effect of measuring dust emission parameters in this work, beyond the simple study in Appendix A.  This is because far IR is needed in order to constrain all the parameters of the \citet{Emission} model which we are able to simulate and fit with \texttt{BAGPIPES}. When only mid-IR is available, as in the cases explored in this work, one should prefer to use constrained IR templates, such as \citep{Dale2001} (one free parameter) or \citep{Chary} or \citep{Salim18} (fully constrained by energy balance). The use of such templates further tightens the determination of $A_V$ and slope because the IR luminosity is better constrained with the longer WISE bands than in the case in which \citet{Emission} is assumed \citep{Salim18, Boquien21}. Unfortunately, this is not currently possible with \texttt{BAGPIPES}, but it will be an interesting study for future work.

These results will be valuable for the upcoming Vera Rubin Observatory Legacy Survey of Space and Time (LSST), which will observe in bands that are very similar to those considered here for DECam ($ugriz$), and that will observe thousands of Supernova host galaxies for which the physical and dust properties will be analyzed. We conclude that complementing LSST data with IR observations to the appropriate depths for the LSST galaxies will be crucial to study their dust properties, and to derive more accurate SFR and stellar mass estimates from photometry alone.

\appendix
\section{Dust Emission Parameters}
\label{sec:emission}
\begin{figure*}
    \includegraphics[width=1\textwidth]{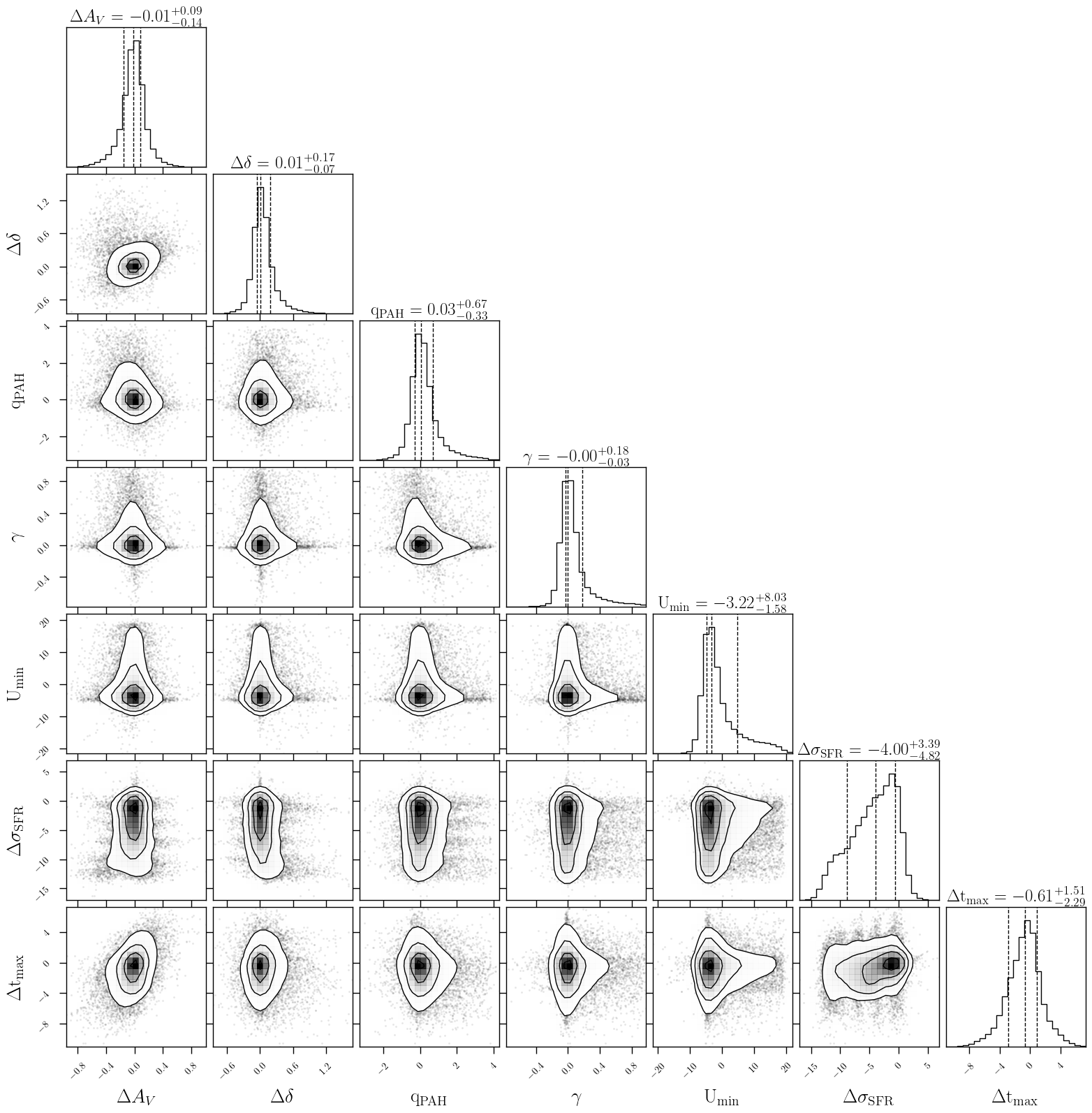}
    \caption{A corner plot of the residuals of parameters of interest, fit using all bands with \texttt{Sim2} model galaxies, with the inclusion of dust emission parameters. This plot demonstrates the difficulty in recovering $U_{min}$.}
    \label{EmissionCorner}
\end{figure*}

Similar to the bump, dust emission is a factor we do not consider for most of our analysis, but we pay special attention to it here. 
 We recover distributions on the dust emission parameters as given in Figure \ref{EmissionCorner}. 

While the emission parameter $q_\mathrm{PAH}$ is relatively well fit, the recovery of $\gamma$ and especially $U_{min}$ is poor. $U_{min}$ shows a highly asymmetric and wide residual distribution, due to the logarithmic prior. The lack of constraining power for this parameter by our data is expected as changes to this parameter would mostly only be covered by one band, W4, and be more clearly observable at longer wavelengths. For what concerns $\gamma$, variations to this parameter also only tend to affect a narrow and red wavelength range, which may fall between two of the W2-W3-W4 bands, and hence also hard to constrain. On the other hand, $q_\mathrm{PAH}$ are bettered covered by the WISE bands, hence they can be decently recovered. In addition to this, the inclusion of the dust emission parameters seems to worsen the recovery of other parameters, giving more scatter in the residuals for all other parameters of interest, see Figure \ref{DustTypes}.

While in most of this paper analysis we have not simulated nor modeled in the SED fitting the IR emission, which we only take into account in this subsection, we not that the modeling in the fitting only matters if one includes FIR bands such as W4. As it is clear from Fig. \ref{AvDeltaContour} the inclusion of W1 and W2 is sufficient, and more important, in the determination of the dust attenuation law, hence our conclusions on the attenuation law remain unchanged by the inclusion of emission at longer wavelengths.

\section*{Data Availability}

The data used in this work can be shared upon reasonable request to the authors. 

\section*{Acknowledgements}
 We thank Adam Carnall for making BAGPIPES public and for their help with the code. Antonella Palmese acknowledges support for this work was provided by NASA through the NASA Hubble Fellowship grant HST-HF2-51488.001-A awarded by the Space Telescope Science Institute, which is operated by Association of Universities for Research in Astronomy, Inc., for NASA, under contract NAS5-26555.

\bibliographystyle{mnras}
\bibliography{references}
\end{document}